\documentclass[aps,prd,amsmath,amssymb,showpacs]{revtex4}
\usepackage{graphicx}
\usepackage{dcolumn}
\usepackage{morefloats}
\usepackage{color}
\usepackage{slashed}
\usepackage{bbm}
\usepackage{subfigure}

\begin{document}
\title{Pseudoscalar charmonium pair interactions via the Pomeron exchange mechanism}

\author{Chang Gong$^1$\footnote{{\it E-mail address:} gongchang@ihep.ac.cn}, Meng-Chuan Du$^{1}$\footnote{{\it E-mail address:} dumc@ihep.ac.cn}, 
Qiang Zhao$^{1,2}$\footnote{{\it E-mail address:} zhaoq@ihep.ac.cn}
}
\affiliation{$^1$ Institute of High Energy Physics,\\
         Chinese Academy of Sciences, Beijing 100049, China}

\affiliation{$^2$ University of Chinese Academy of
Sciences, Beijing 100049, China}

\begin{abstract}

Following the observation of the fully-heavy tetraquark candidates $X(6900)$ and possible $X(6300)$ in the di-$J/\psi$ spectrum by the LHCb Collaboration, we investigate the near-threshold dynamics of the pseudoscalar quarkonium pairs in a coupled-channel approach in the di-$\eta_c$ channel. We show that the Pomeron exchange mechanism should be a general dynamics in the near-threshold heavy quarkonium pair interactions. In the di-$\eta_c$ channel, a coupled-channel system involving the di-$\eta_c$, $\eta_c$-$\eta_c(2S)$ and di-$\eta_c(2S)$ interactions can be established. Their $S$-wave interactions near threshold via the Pomeron exchange potential can produce near-threshold resonance poles similar to the di-$J/\psi$, $J/\psi$-$\psi(2S)$ and di-$\psi(2S)$ coupled channels. Interestingly, we find here that the pole positions are more shifted to the corresponding thresholds and turn out to be different from those seen in the di-$J/\psi$ spectrum. Taking into account the suppression of the heavy quark spin flips in the transitions between two vector and two pseudoscalar heavy quarkonium systems, the enhancements in the di-$\eta_c$ spectrum should be different from those seen in the di-$J/\psi$ channel. Experimental study of the di-$\eta_c$ channel should be useful for disentangling the nature of the fully-heavy tetraquark systems.  

\end{abstract}

\maketitle

\section{Introduction}

The recent observation of the narrow enhancement $X(6900)$ and a possible broad structure $X(6300)$ in the di-$J/\psi$ spectrum by the LHCb Collaboration~\cite{Aaij:2020fnh} has provided a strong evidence for exotic hadrons made of four heavy quarks, which are beyond the conventional quark model scenario. This observation immediate provokes tremendous interests in its nature and various proposals for its structure are made in the literature. 

The fully-heavy tetraquark systems, such as $cc\bar{c}\bar{c}$, $bb\bar{b}\bar{b}$, $cb\bar{c}\bar{b}$, are peculiarly interesting since the leading-order quark (anti-quark) interactions do not involve the light quark degrees of freedom. One would anticipate that the short-distance color interactions play a dominant role and such multiquark systems should be genuine compact color-singlet tetraquark states rather than loosely bound hadronic molecules where in many cases long-distance light hadron exchanges are expected to provide the binding mechanism.

Early studies of the fully-heavy tetraquark systems can be found in the literature~\cite{Ader:1981db,Iwasaki:1975pv,Zouzou:1986qh,Heller:1985cb,Lloyd:2003yc,Barnea:2006sd}. With the recent experimental progresses theoretical studies of the fully-heavy tetraquark states are carried out based on different treatments for the heavy quark (antiquark) interactions~\cite{Vijande:2009kj,Berezhnoy:2011xn,Wang:2017jtz,Karliner:2016zzc,Bai:2016int,Anwar:2017toa,Esposito:2018cwh,Chen:2016jxd,
Wu:2016vtq,Hughes:2017xie,Richard:2018yrm,Debastiani:2017msn,Wang:2018poa,Richard:2017vry,Deng:2020iqw,Ohlsson,
Wang:2019rdo,Bedolla:2019zwg,Chen:2020lgj,Chen:2018cqz,Liu:2019zuc}. These analyses seem to have different conclusions on the property of the fully-heavy tetraquark states. In Refs.~\cite{Wang:2017jtz,Karliner:2016zzc,Berezhnoy:2011xn,Bai:2016int,Anwar:2017toa,Esposito:2018cwh,Debastiani:2017msn,Wang:2018poa} the fully-heavy tetraquark states are predicted to have masses below the heavy charmonium or bottomonium pair thresholds. In such a scenario these states will keep stable since direct decays into heavy quarkonium pairs via quark rearrangements are forbidden. In contrast, other studies show that the fully-heavy tetraquark states may have masses above the thresholds of the corresponding heavy quarkonium pairs~\cite{Ader:1981db,Lloyd:2003yc,Wu:2016vtq,Hughes:2017xie,Richard:2018yrm,Richard:2017vry,Liu:2019zuc,Deng:2020iqw,Wang:2019rdo,Chen:2016jxd,Chen:2018cqz,liu:2020eha}. As discussed in Ref.~\cite{Liu:2019zuc,liu:2020eha} the controversy arises from the different treatments of the linear confinement potential. It shows that an explicit inclusion of this potential will significantly raise the eigenvalues of the ground states and lead to resonance solutions for these fully-heavy tetraquark states.

Following the observation of $X(6900)$ various interpretations are proposed in order to understand the nature of $X(6900)$. In particular, in the potential models $X(6900)$ is far above the ground state and can be assigned as either the first radial excitation states of $0^{++}/2^{++}$ or the first orbital excitation state of $0^{-+}/1^{-+}$~\cite{liu:2020eha,Wang:2020gmd,Garcilazo:2020acl,Giron:2020wpx,Sonnenschein:2020nwn,Maiani:2020pur,Richard:2020hdw,Chao:2020dml,Maciula:2020wri,Karliner:2020dta,Ma:2020kwb,Cao:2020gul,Zhu:2020snb,Guo:2020pvt,Zhu:2020xni,Weng:2020jao,Majarshin:2021hex,Wang:2021kfv,Richard:2021nvn,Li:2021ygk,Mutuk:2021hmi,Yang:2021hrb,Ke:2021iyh,Faustov:2021hjs,Huang:2020dci}. QCD sum rules are also employed to provide information for its tetraquark nature~\cite{Wang:2020dlo,Wan:2020fsk,Yang:2020wkh}. However, the experimental evidence seems to raise more questions on the fully-heavy tetraquarks instead of solving problems. One immediate question is that why there are only one narrow structure is seen, given that the potential model would predict a large number of states even for the same quantum numbers.  In Ref.~\cite{liu:2020eha} some configuration cancellation effects can be identified for the radial excitation of $0^{++}$ decays into di-$J/\psi$. But we still lack quantitative results for understanding the potential quark model spectra, and it is non-trivial to reach a direct answer. In Refs.~\cite{Dong:2020nwy,Dong:2021lkh,Zhuang:2021pci,Liang:2021fzr} coupled-channel approaches with effective short-distance potentials are applied to investigate the vector charmonium pair interactions. With unitarization resonance pole structures are identified and the line shape of the di-$J/\psi$ spectrum can be described. The coupled-channel approach can indeed evade the problem raised by the potential quark model, namely, the number of states observed in experiment is far less than that predicted by the potential quark model. However, it is unclear what mechanism provides the short-distance effective potential in the coupled-channel approaches. 

In Ref.~\cite{Gong:2020bmg} we propose that the Pomeron exchange plays a unique role in heavy-quarkonium scatterings. It provides a soft scale in the interactions between two vector charmonia, i.e. $J/\psi$-$J/\psi$, and $J/\psi-\psi(2S)$, near threshold. With unitarization it shows that pole structures can arise from the Pomeron exchange and coupled-channel effects between the  di-$J/\psi$, $J/\psi-\psi(2S)$, and di-$\psi(2S)$ scatterings. For the heavy meson scatterings, the Pomeron exchange has strong couplings near threshold. It provides an origin of the short-distance dynamics for the $X(6900)$ structure in the di-$J/\psi$ spectrum, and predicts a general feature for double heavy quarkonium scatterings near threshold. 

In this work we extend the Pomeron exchange dynamics to the di-$\eta_c$ and $\eta_c-\eta_c(2S)$ coupled channel, and explore its role played in different heavy-quarkonium scattering processes. As discussed in Ref.~\cite{Gong:2020bmg} a special feature with the Pomeron exchange in the heavy-quarkonium scatterings is that the Pomeron exchange can occur via both the $t$ and $u$ channel. Interferences between the $t$ and $u$-channel amplitudes are strong near threshold and they will die out with the increase of the center of mass (c.m.) energy. For the di-$\eta_c$ and $\eta_c-\eta_c(2S)$ scatterings, the Pomeron-$\eta_c$ ($\eta_c(2S)$) vertex has different coupling structure from the Pomeron-$J/\psi$ ($\psi(2S)$). But in the near-threshold region the dominant terms have similar behaviour. It is also interesting to note that the relative $S$-wave di-$\eta_c$ or $\eta_c-\eta_c(2S)$ couple to $J^{PC}=0^{++}$. Meanwhile, one would expect little mixing between di-$J/\psi$ and di-$\eta_c$ in the state of $0^{++}$. The reason is that the transition between $J/\psi$ and $\eta_c$ is an $M1$-type transition which will be suppressed by $p/m_c$. Here, $p$ is the typical three-momentum of the charm quark in the near-threshold kinematic region. Thus, structures observed in the di-$\eta_c$ spectrum should be distinguished from the structures seen in the di-$J/\psi$ spectrum.

In the next Section we briefly introduce the Pomeron exchange dynamics and provide the coupled-channel formalism for the di-$\eta_c$ and $\eta_c-\eta_c(2S)$ scatterings. In Sec.~\ref{Results} the numerical results and discussions are presented. A brief summary is given in the last Section.

\section{Formalism} \label{Formalism}

\subsection{Pomeron exchange mechanism}

Pomeron as an effective degrees of freedom for the $t$-channel multi-soft-gluon exchanges has been extensively studied in the literature~\cite{Donnachie:1984xq,Donnachie:1987pu}. It has been successfully applied to high-energy processes to account for the diffractive  behaviors in hadron collisions and vector meson photo or electroproductions on the nucleon~\cite{Pichowsky:1996jx,Pichowsky:1996tn,Laget:1994ba,Zhao:1999af}. As noted in Ref.~\cite{Gong:2020bmg}, the Pomeron exchange is different from the $t$-channel pole contributions. It behaves rather like a positive charge conjugation isoscalar photon with $J^{PC}=1^{-+}$, and does not have a pole in the positive angular momentum complex plane. This makes it different from the $t$-channel light meson exchanges although sometimes the soft-gluon exchanges may give rise to effective light meson exchanges due to quark-hadron duality argument~\cite{Dolen:1967jr}. For the heavy-quarkonium scatterings near threshold, the absence of the Regge trajectories for light-meson exchanges implies the leading role played by the Pomeron trajectory.

The asymptotic soft-gluon dynamics can be described by the Pomeron trajectory in the Regge theory, i.e.
\begin{equation}
i \mathcal{G}(s,t)=(\alpha' s)^{\alpha(t)-1} \ ,
\end{equation}
where $\alpha(t)\equiv 1+\epsilon'+\alpha' t$ with $\epsilon'=0.08$ a small positive quantity indicating the dominance of the $C=+1$ Pomeron exchange in the $t$ channel, and $\alpha'=0.25$ GeV$^{-2}$ as commonly adopted value.

\begin{figure}[h]
\centering
\subfigure[ ~$t$ channel process]{
\includegraphics[width=6.5cm]{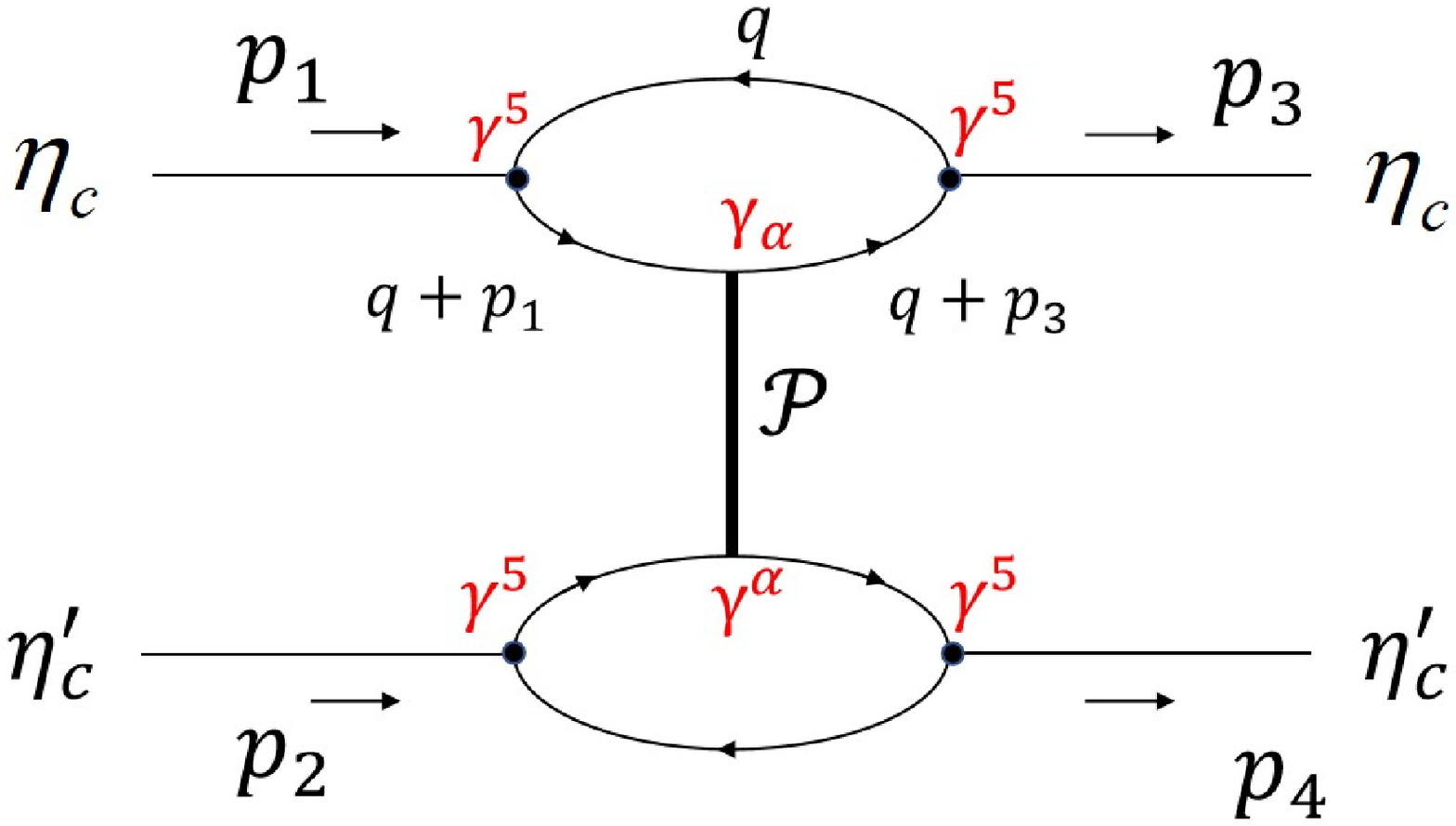}
}
\quad
\subfigure[ ~$u$ channel process]{
\includegraphics[width=6.5cm]{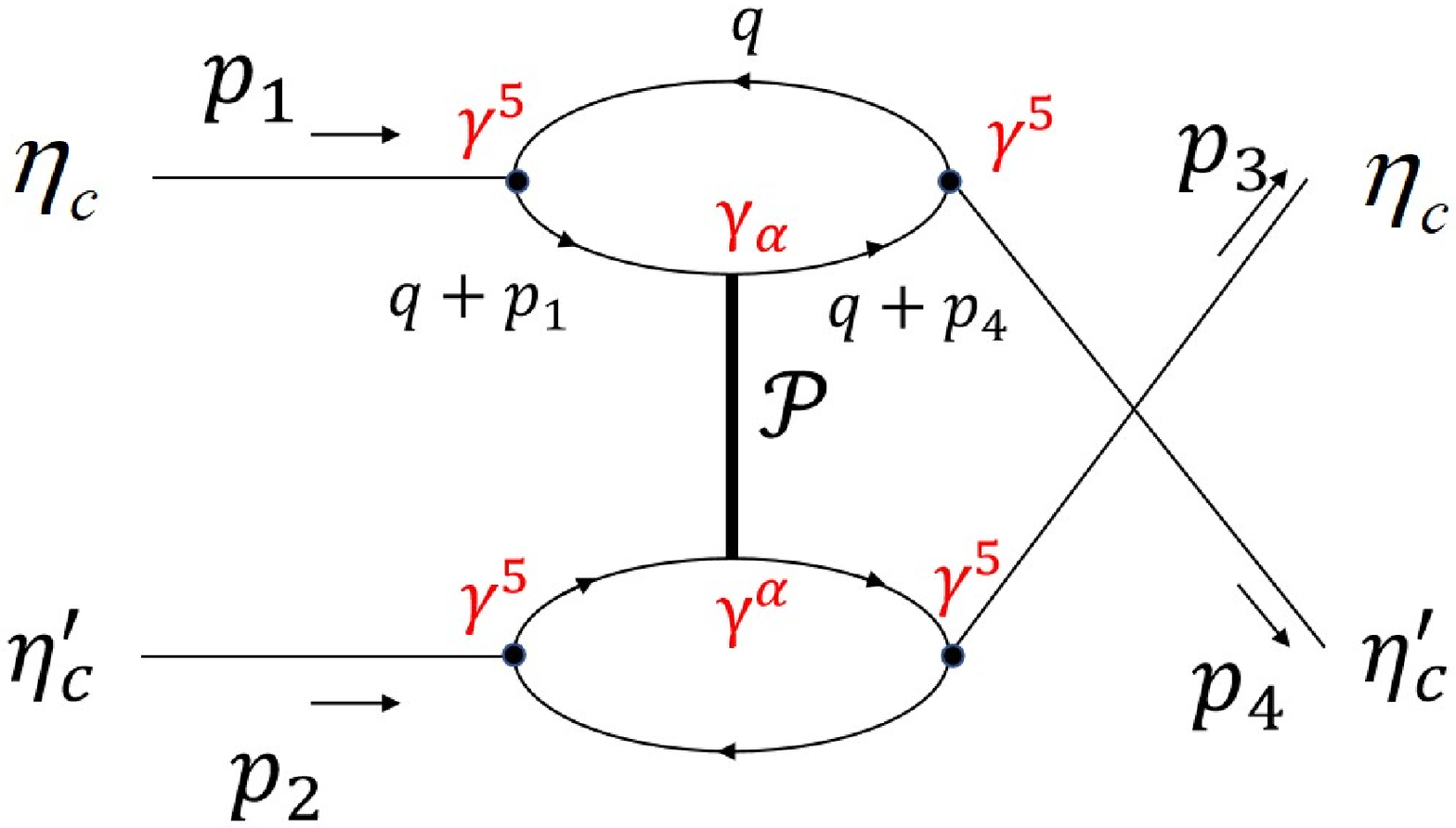}
}
\caption{Illustrative diagrams for (a) $t$ channel and (b) $u$ channel Pomeron exchange processes in $\eta_{c} \eta_{c}(2S) \to \eta_{c} \eta_{c}(2S) $. For $\eta_c\eta_c\to \eta_c\eta_c$ Bose symmetry will introduce a factor of 2 subtraction to the cross section.}
\label{fig1}
\end{figure}

The general form of the interaction between the Pomeron and the constituent quarks can be expressed as 
\begin{equation}
\beta_c^2\bar{Q}_i\gamma_\alpha Q_i\bar{Q}_j\gamma^\alpha Q_j  \mathcal{G}(s,t) \ ,
\end{equation}
where $Q_i$ and $Q_j$ are the constituent quarks (anti-quarks) in the initial hadrons which interact with each other by the Pomeron exchange; $\beta_{c}=1.17 \ \text{GeV}^{-1}$ is the coupling strength between the Pomeron and the charm quark in the charmonium states. We adopt the pseudoscalar scalar coupling for the $\eta_c Q\bar{Q}$ vertex, $\bar{Q}\gamma_5 Q \eta_c$. Then, the quark loop amplitude can be factorized as~\cite{Laget:1994ba},
\begin{equation}\label{trans-amp-1}
\tilde{t}^{\alpha}(\eta_{c})={\cal F}_{\eta_{c}}(t)\  \Gamma^{\alpha},
\end{equation}
where ${\cal F}_{\eta_{c}}(t)$ is the form factor of the corresponding hadron which arises from the transition between the initial and final hadrons via the Pomeron exchange with momentum transfer $t$; $\Gamma^{\alpha}$ is the spin structure for the quark interactions with the Pomeron and external hadrons in one of the quark loops in Fig.~\ref{fig1}, and has the following form:
\begin{equation}
\Gamma^{\alpha} \propto \text{Tr} \ [\gamma^{5}  (\slashed{q}+m_{c})   \gamma^{\alpha}  (\slashed{q} +\slashed{p}_{1} +m_{c}) \gamma^{5}  (\slashed{q} +\slashed{p}_{3} +m_{c})],
\end{equation}
where $m_{c}$ is the constituent charm quark mass of the interacting quark and $q$ is the four-vector momentum introduced into the loop function in Fig.~\ref{fig1}. Note that the on-shell

For the Pomeron exchange transitions between two heavy flavor hadrons near threshold, the Regge limit can be satisfied immediately, i.e. $s>> |t|$. It allows the on-shell approximation~\cite{Pichowsky:1996jx}, and function $\Gamma^{\alpha}$ can be reduced to
\begin{eqnarray}
\Gamma^{\alpha}&=&\frac{1}{M_{\eta_c}^{2}}[(4 m_{c}^{2}+ p_{3}^{2})p_{1}^{\alpha}+\frac{1}{2}(4 m_{c}^{2}- p_{3}^{2})p_{3}^{\alpha}] \ ,
\end{eqnarray}
where $m_{c}={M_{\eta_{c}}}/{2}$ has been adopted. Applying the formalism to both the $t$ and $u$ channel, the $\eta_{c}-\eta_{c}$ scattering amplitude can be written as,
\begin{eqnarray}\label{t-u-amp}
T_{t}^{\mathcal{P}}&=& ({\cal F}_{\eta_{c}}(t))^{2} \times \mathcal{G}(s,t) \times \frac{1}{m_{\eta_c}^{2}}[(4 m_{c}^{2}+ p_{3}^{2})p_{1}^{\alpha}+\frac{1}{2}(4 m_{c}^{2}- p_{3}^{2})p_{3}^{\alpha}] \times \frac{1}{m_{\eta_c}^{2}}[(4 m_{c}^{2}+ p_{4}^{2})p_{2}^{\alpha}+\frac{1}{2}(4 m_{c}^{2}- p_{4}^{2})p_{4}^{\alpha}]   \\
T_{u}^{\mathcal{P}}&=& ({\cal F}_{\eta_{c}}(u))^{2} \times \mathcal{G}(s,t) \times\frac{1}{m_{\eta_c}^{2}}[(4 m_{c}^{2}+ p_{4}^{2})p_{1}^{\alpha}+\frac{1}{2}(4 m_{c}^{2}- p_{4}^{2})p_{4}^{\alpha}] \times \frac{1}{m_{\eta_c}^{2}}[(4 m_{c}^{2}+ p_{3}^{2})p_{2}^{\alpha}+\frac{1}{2}(4 m_{c}^{2}- p_{3}^{2})p_{3}^{\alpha}]
\end{eqnarray}
where $p_{1}$ and $p_{2}$ are the four-momentum of the initial pseudoscalar meson. 
In the limit of $|t|\to 0$, function $\Gamma^{\alpha}$ can be further simplified as $\Gamma^{\alpha} \approx  2 p_{1}^{\alpha}$. This behaviour actually favors the heavy-flavor hadron interactions near threshold via the Pomeron exchange where, on the one hand, the amplitude will be enhanced by the quark mass, and on the other hand, the form factor suppression is minimized. In contrast, for light hadron scatterings with light flavors, the Pomeron exchange contribution will be suppressed by the quark masses near threshold. In such a sense, the Pomeron exchange actually plays a unique role in heavy quarkonium pair scatterings near threshold and has not been noticed before.

It should be noted that the $t$ and $u$-channel in Eq.~(\ref{t-u-amp}) will equal to each other in the scatterings of two identical mesons, e.g. $J/\psi J/\psi\to J/\psi J/\psi$ and $\eta_c\eta_c\to\eta_c\eta_c$. Thus, a factor of 2 should be subtracted in the cross section due to  Bose symmetry in these scatterings. 

In Eq.~(\ref{trans-amp-1}) function ${\cal F}_{\eta_{c}}(t)$ is the form factor which contains the dynamics of the Pomeron-meson interaction. It describes the momentum transfer dependence of the coupling strength due to the Pomeron exchange. We adopt a commonly used form~\cite{Donnachie:2002en} as follows:
\begin{equation}
{\cal F}_{\eta_{c}}(t)=(2 \beta_{c}) \ \exp(\frac{t}{2\lambda_{\eta_{c}}^2}),
\end{equation}
where $\beta_{c}=1.17 \ \text{GeV}^{-1}$ is the coupling between Pomeron and the $c$ quark in the meson. It is determined by fitting the experimental data for the $J/\psi$ photoproduction~\cite{Chekanov:2002xi}. The parameter $\lambda_{\eta_{c}}=1.2$ GeV is a typical energy scale reflecting the Pomeron-valence-quark interaction range.

The effective potential $V(s,t)$ at leading order is obtained by the sum of the $t$ and $u$-channel Pomeron exchange amplitudes. Note that the explicit $t$-dependence appears in the Pomeron exchange potential. We only focus on the $S$-wave amplitudes in the pseudoscalar charmonium scatterings, and the quantum number of the $S$-wave couplings between the two pseudoscalar charmonia can only be $0^{++}$. By integrating the angular part of the Pomeron exchange potential, we define an equivalent separable potential following the unitarization scheme of Ref.~\cite{Roca:2005nm},
\begin{equation}\label{separa-pot}
V(s)=\frac{1}{2} \int V(s,t) d (\cos{\theta}).
\end{equation}
This approximation is justified for the relative $S$-wave couplings near threshold. It is also a reasonable approximation for the heavy quarkonium interaction systems near threshold with $s>>|t|$ and $s>> |u|$ in Fig.~\ref{fig1}.

We apply the Bethe-Salpeter equation on the separable potential $V(s)$ for the heavy quarkonium system consisting of $H_1$ and $H_2$ and extract the general form of the $T$-matrix,
\begin{equation}
T(s)=\frac{V(s)}{1-\tilde{G}(s) V(s)} \ ,
\label{T-matrix}
\end{equation}
where the loop function $\tilde{G}(s)$ is
\begin{equation}
\tilde{G}(s)= \int \frac{d^{4}q}{(2 \pi)^{4}} \frac{i^{2}\exp{(-2 \vec{q}^2 / \Lambda^{2})}}{[q^{2}-M_{H_1}^{2}+i \epsilon][(P-q)^{2}-M_{H_2}^{2}+i \epsilon]} \ ,
\end{equation}
with $M_{H_1}$ and $M_{H_2}$ denoting the masses of the two heavy charmonia, respectively. 
Note that the integral is regularized by a cut-off function $\exp{(-2 \vec{q}^2 / \Lambda^{2})}$ as long as the internal charmonium states go off-shell. The analytical expression of the integral is~\cite{Guo:2014iya,Cao:2017lui},
\begin{equation}
\tilde{G}(s)=\frac{i}{4M_{H_1}M_{H_2}}\left[-\frac{\mu \Lambda}{(2\pi)^{3/2}}+\frac{\mu k}{2\pi}\exp{(-2k^{2}/\Lambda^{2})}[erfi(\frac{\sqrt{2}k}{\Lambda})-i]\right],
\end{equation}
where $k\equiv\sqrt{2\mu(\sqrt{s}-M_{H_1}-M_{H_2})}$; $\Lambda= 0.7$ GeV is the form factor parameter corresponding to the typical size of the heavy quarkonia; and $\mu\equiv \frac{M_{H_1}M_{M_2}}{M_{H_1}+M_{H_2}}$ is the reduced mass; and $erfi(\frac{\sqrt{2}k}{\Lambda})$ is the imaginary error function.

\subsection{Coupled-channel approach with the effective potential}\label{Coupled-Channel Model}

Similar to the study of the coupled channels of the di-$J/\psi$, $J/\psi$-$\psi(2S)$ and di-$\psi(2S)$ system~\cite{Gong:2020bmg}, the coupled channels of the di-$\eta_{c}$, $\eta_{c}$-$\eta_{c}(2S)$ and di-$\eta_{c}(2S)$ can be described by a $3\times 3$ potential, i.e. 
\begin{equation}
V(s)
=
\left(
\begin{array}{ccc}
    V_{11} & V_{12}  & V_{13} \\
    V_{21} & V_{22}  & V_{23} \\
    V_{31} & V_{32}  & V_{33} 
\end{array}
\right),
\end{equation}
where $V_{ij}$ denotes the Pomeron exchange potentials including both the $t$ and $u$ channels for each process.
The loop integral function $G$ for the di-$\eta_{c}$, $\eta_{c}$-$\eta_{c}(2S)$ and di-$\eta_{c}(2S)$ coupled channels is written as
\begin{equation}
G(s)
=
\left(
\begin{array}{ccc}
    G_{1} & 0  & 0 \\
    0 & G_{2} & 0 \\ 
    0 & 0 & G_{3}
\end{array}
\right).
\end{equation}
To evaluate the coupled-channel contributions at LHCb, the energy spectrum should be an input, which however is unavailable. Thus, we adopt the same prescription of the energy spectrum for the di-$J/\psi$ production as Refs.~\cite{Dong:2020nwy,Gong:2020bmg}, and the transition amplitude (labelled as channel 1) is written as 
\begin{equation}\label{di-etac-amp}
\mathcal{M}_{1}=P(\sqrt{s})(1+\sum {r_{i}G_{i}(s)}T_{i1}(s)),
\end{equation}
with $T_{i1}(s)$ being the element of the $T$-matrix in Eq.~(\ref{T-matrix}). The ratios $r_{i}$ describe the different production strengths for different channels. The function $P(\sqrt{s})$ parametrize out the energy spectrum of the short-distance production as follows:
\begin{equation}
P(\sqrt{s})=\alpha e^{-\beta s} \ ,
\label{distribution}
\end{equation}
where $\beta$ and $\alpha$ are experiment-related quantities. For the di-$\eta_c$ production, the lack of experimental information means that we cannot provide a definite prediction for the interfering pattern caused by these coupled channels in the di-$\eta_c$ spectrum. As a qualitative estimate, we apply $\beta=0.012 \ GeV^{-2}$ and  $\alpha=0.31$ which are fitted in the di-$J/\psi$ production~\cite{Dong:2020nwy,Gong:2020bmg}.

We also need to fix the unknown parameter $r_{i}$ in Eq. \ref{di-etac-amp}, which determines the relative production strengths for different channels at LHCb. However, due to the lack of the experimental information, we adopt the same values for $r_{i}$ as in Ref.~\cite{Gong:2020bmg}. Meanwhile, since it is possible that $r_{i}$ carry complex phases  arising from the production mechanism, e.g. higher resonance channels can feed down to the di-$\eta_{c}$ spectrum via the DPS processes, we will investigate the phase dependence of the transition amplitude in the di-$\eta_c$ spectrum in the numerical calculations.

In our model the di-$\eta_{c}$ spectrum is calculated by
\begin{equation}
\Gamma(s)=\frac{|\vec{p}_{\eta_{c}}|}{8 \pi s} |\mathcal{M}_{1}|^{2},
\end{equation}
where $\vec{p}_{ \eta_{c}}$ is the three-momentum of the final $\eta_{c}$ in the center of mass frame of the initial states.

\section{Numerical results and discussions}\label{Results}

Proceeding to the numerical calculations of the di-$\eta_{c}$ energy spectrum, we first consider the pole structures in the $T$-matrix for single channels. It shows that resonance poles with  $J^{PC}=0^{++}$ can be produced by the Pomeron exchange potential for the pseudoscalar charmonium pairs. 

\begin{figure}[h]
\centering
\includegraphics[height=5cm,width=8cm]{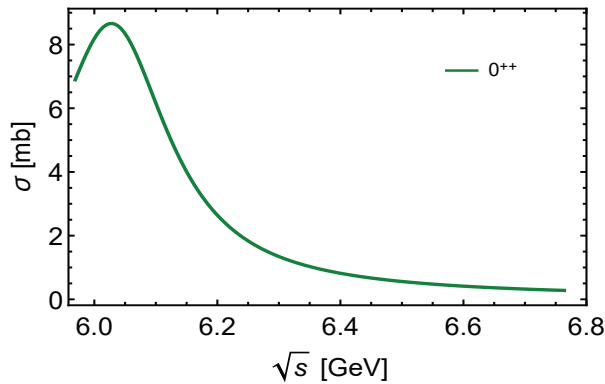}
\caption{Single channel scattering cross sections for $\eta_{c} \ \eta_{c} \to \eta_{c} \  \eta_{c}$ via the Pomeron exchange. }
\label{di-etac}
\end{figure}

\begin{figure}[h]
\centering
\includegraphics[height=5cm,width=8cm]{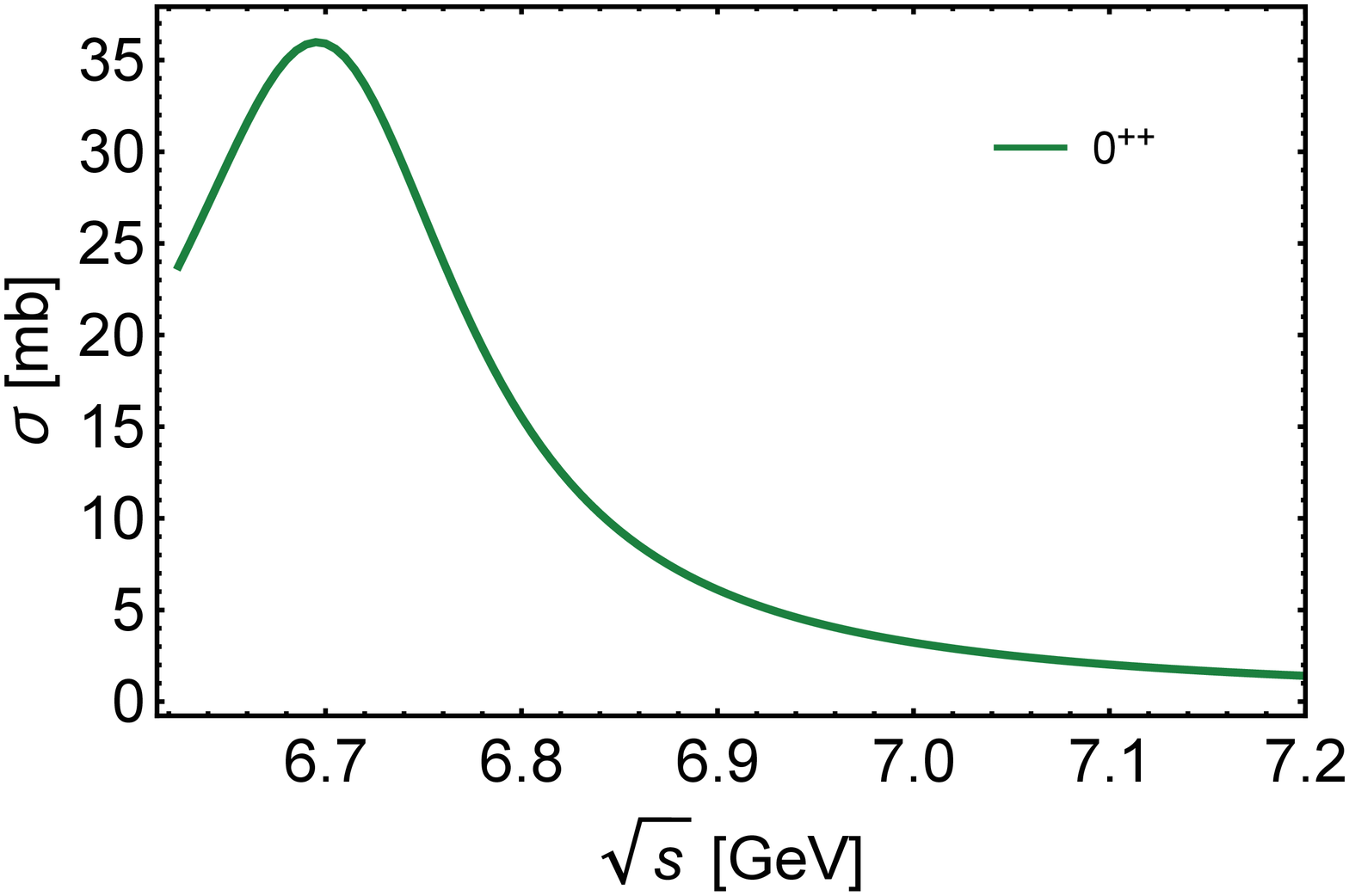}
\caption{Single channel scattering cross sections for $\eta_{c} \ \eta_{c}(2S) \to \eta_{c}  \ \eta_{c}(2S)$ via the Pomeron exchange. }
\label{etac-etac(2S)}
\end{figure}

In Figs.~\ref{di-etac} and ~\ref{etac-etac(2S)}, we illustrate the spectra for single channel scatterings, i.e.  $\eta_{c} \ \eta_{c} \to \eta_{c} \  \eta_{c}$ and $\eta_{c} \ \eta_{c}(2S) \to \eta_{c}  \ \eta_{c}(2S)$, respectively. Enhancement structures can be produced above their corresponding thresholds and we obtain two resonance poles which are located at $(6031-i 110 )$ and $(6703-i 90)$ MeV, respectively, on the second Riemann sheet of the single channels. The structures turn out to be different from those in the di-$J/\psi$ and $J/\psi$-$\psi(2S)$ single channel scatterings~\cite{Gong:2020bmg}. It shows that the peaks are more shifted towards the thresholds here than in the system of the vector pairs.  Such a difference is due to the different coupling vertices for the pseudoscalar and vector mesons to the constituent quark and antiquark pair. 

It is unlikely that the $J^{PC}=0^{++}$ structures in the di-$\eta_c$ spectrum are the same as those seen in the di-$J/\psi$ spectrum. Taking into account that the transition of $J/\psi\to \eta_c$ involves the spin flip of the heavy quark, the transition of $J/\psi J/\psi\to \eta_c\eta_c$ should be suppressed in the heavy quark symmetry limit. In other words, it suggests that the Pomeron exchange mechanism actually predicts different near-threshold resonance structures with $J^{PC}=0^{++}$ in the di-$J/\psi$ and di-$\eta_c$ channels. This makes the measurement of the di-$\eta_c$ channel interesting for disentangling the nature of $X(6900)$ and the possible near-threshold state $X(6300)$ in the di-$J/\psi$ spectrum. It may also provide a direct test of the Pomeron exchange mechanism if a lower mass near-threshold enhancements can be identified.

For the coupled channel, we further simplify this study by reducing the $3\times 3$ coupled-channel problem into a $2\times 2$ coupled-channel, i.e. with the effective potential
\begin{equation}
V(s)
=
\left(
\begin{array}{cc}
    V_{11} & V_{12}   \\
    V_{21} & V_{22}   \\ 
\end{array}
\right),
\end{equation}
and the loop integral function
\begin{equation}
G(s)
=
\left(
\begin{array}{cc}
    G_{1} & 0  \\
    0 & G_{2}  
\end{array}
\right).
\end{equation}
This reduction is because we only focus on the di-$\eta_c$ spectrum of which the threshold is far away from the di-$\eta_c(2S)$ threshold. As a result, the di-$\eta_c(2S)$ interaction contributions to the di-$\eta_c$ channel will be signficantly suppressed by the Pomeron-meson form factor if the initial and final-state mesons have large mass differences. We note that the di-$\eta_c(2S)$ interaction will become significant in the $\eta_c$-$\eta_c(2S)$ channel as the coupled-channel effects. The other reason for this approximation is, as mentioned earlier, due to the lack of experimental information on the di-$\eta_c$ production. Therefore, focussing on the di-$\eta_c$ channel near threshold will allow us to focus on the mechanisms which are predominant in this channel.

By searching for the pole structures in the coupled-channel $T$-matrix (Eq.~\ref{T-matrix}), we identify two resonance poles, $(6041-i 170)$ MeV and $(6711-i 141)$ MeV, which are located at the second and fourth Riemann sheet, respectively. It is similar to the case of the $S$-wave di-$J/\psi$ and $J/\psi-\psi(2S)$ coupled channel~\cite{Gong:2020bmg}, and reflects some general features arising from the Pomeron exchange mechanism. Comparing with the poles, $(6278-i 80)$ MeV and $(6860- i 74)$ MeV, found in  the $S$-wave di-$J/\psi$ and $J/\psi-\psi(2S)$ coupled channel interactions, we still see that the poles created by the di-$\eta_c$ and $\eta_c$-$\eta_c(2S)$ interactions have lower masses and are shifted more towards the corresponding thresholds than in the coupled channels of  the di-$J/\psi$ and $J/\psi-\psi(2S)$.

For the di-$\eta_c$ spectrum in its production at LHC, the experimental information is still lacking. We thus adopt the similar production relation as in the di-$J/\psi$ production~\cite{Gong:2020bmg} to investigate the di-$\eta_c$ spectrum.  Considering that $\eta_c$ and $J/\psi$ both are the ground state charmonia, and both $\eta_c(2S)$ and $\psi(2S)$ are well defined first radial excitation states, we fix the relative production rate but leave the phase angle as a free parameter in the numerical calculations, i.e. $r_{1}:r_{2}=1:2 e^{i \delta}$. The dependence of the coupled-channel spectrum  on the phase angle $\delta$ will be illustrated by the line shapes at different values for $\delta$.

\begin{figure}[h]
\centering
\subfigure[ \ $\delta=-\pi/4$]{
\includegraphics[width=3.5cm]{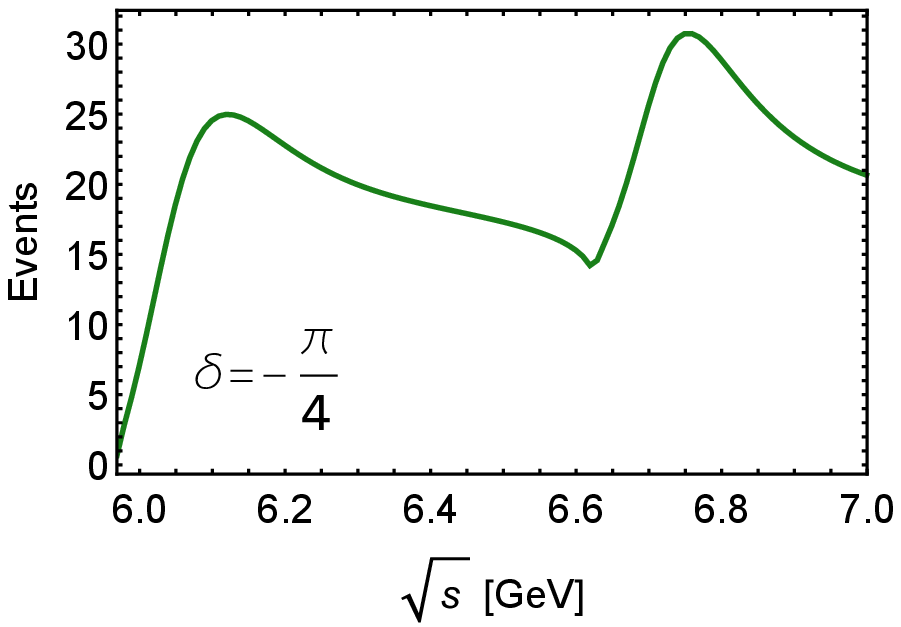}
}
\quad
\subfigure[ \ $\delta=-\pi/2$]{
\includegraphics[width=3.5cm]{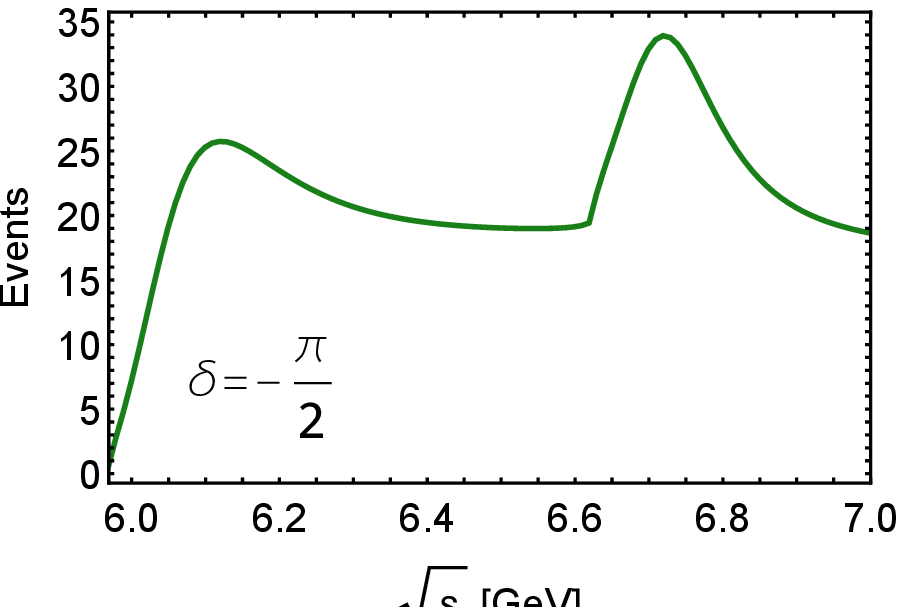}} 
\quad
\subfigure[ \ $\delta=-3 \pi/4$]{
\includegraphics[width=3.5cm]{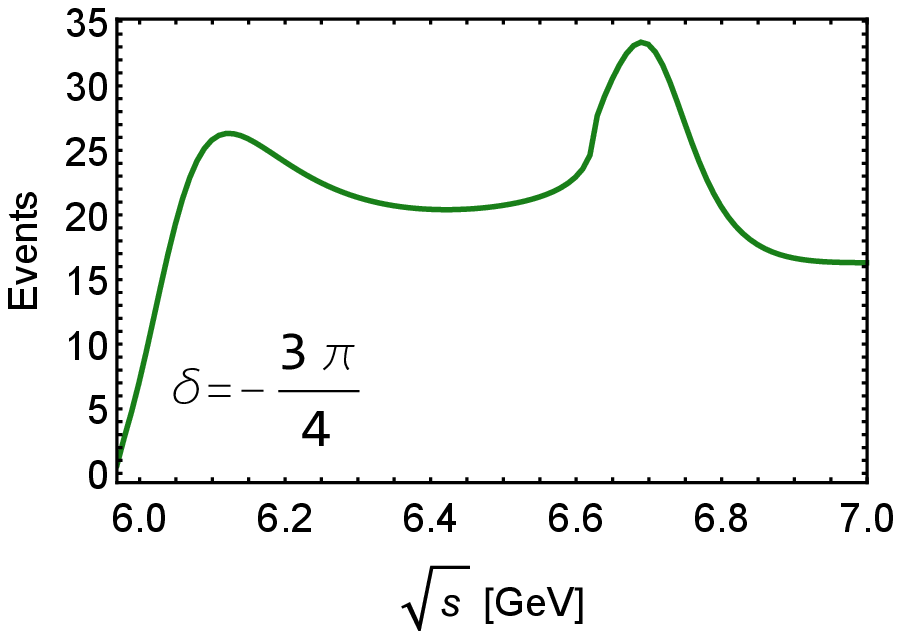}
}
\quad
\subfigure[ \ $\delta=-\pi$]{
\includegraphics[width=3.5cm]{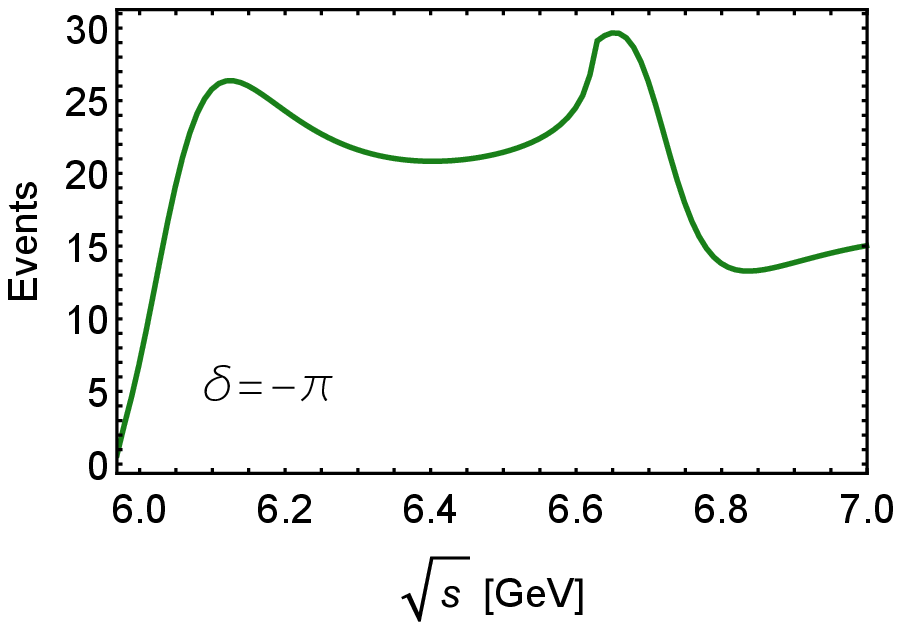}
}
\\
\subfigure[ \ $\delta= \pi/4$]{
\includegraphics[width=3.5cm]{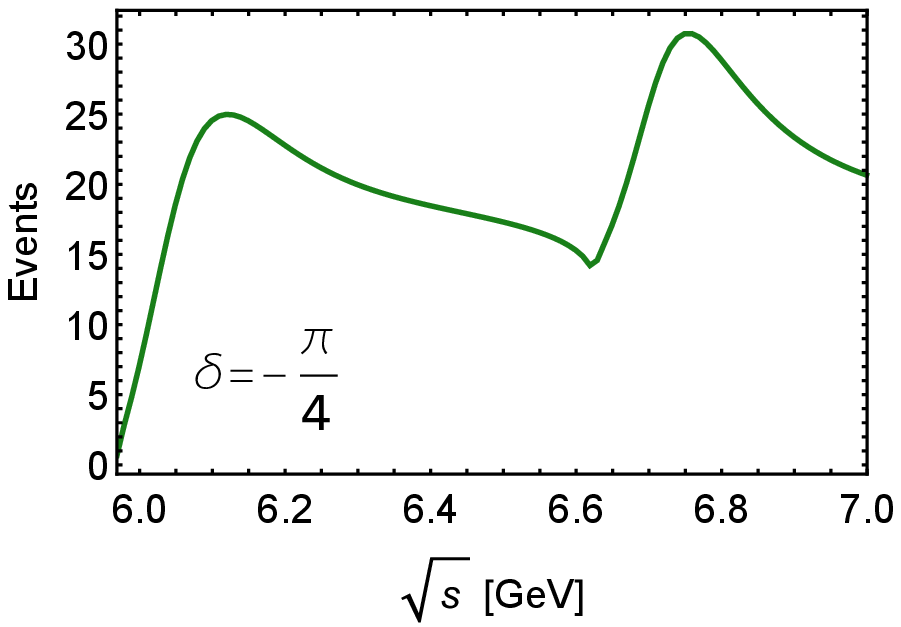}
}
\quad
\subfigure[ \ $\delta= \pi/2$]{
\includegraphics[width=3.5cm]{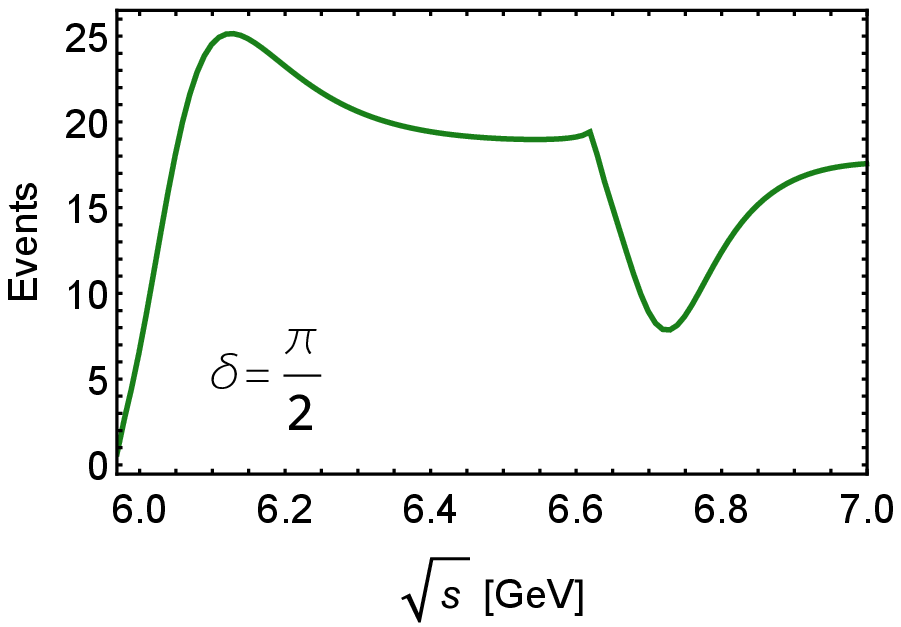}
}
\quad
\subfigure[ \ $\delta=3\pi/4$]{
\includegraphics[width=3.5cm]{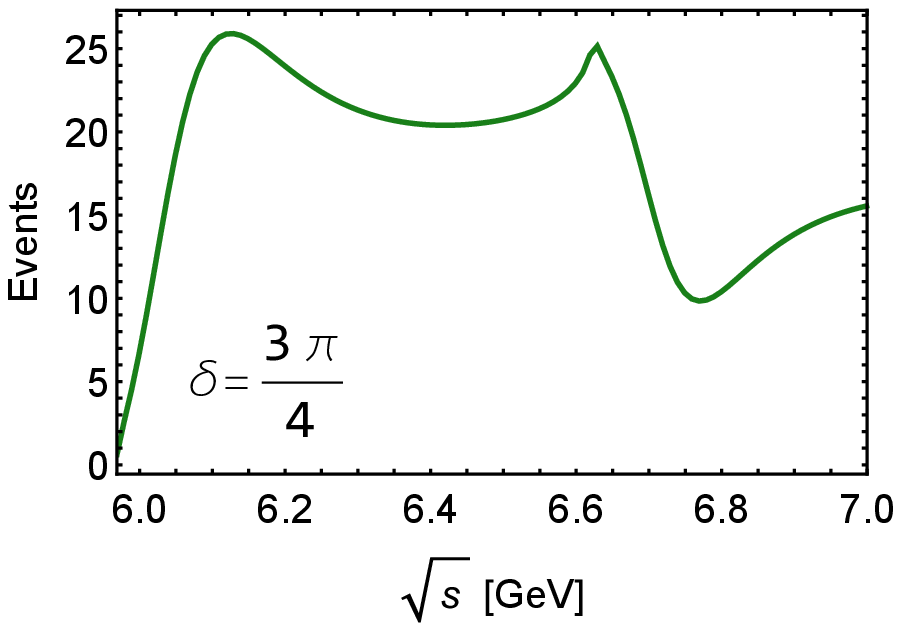}
}
\quad
\subfigure[ \ $\delta= 0$]{
\includegraphics[width=3.5cm]{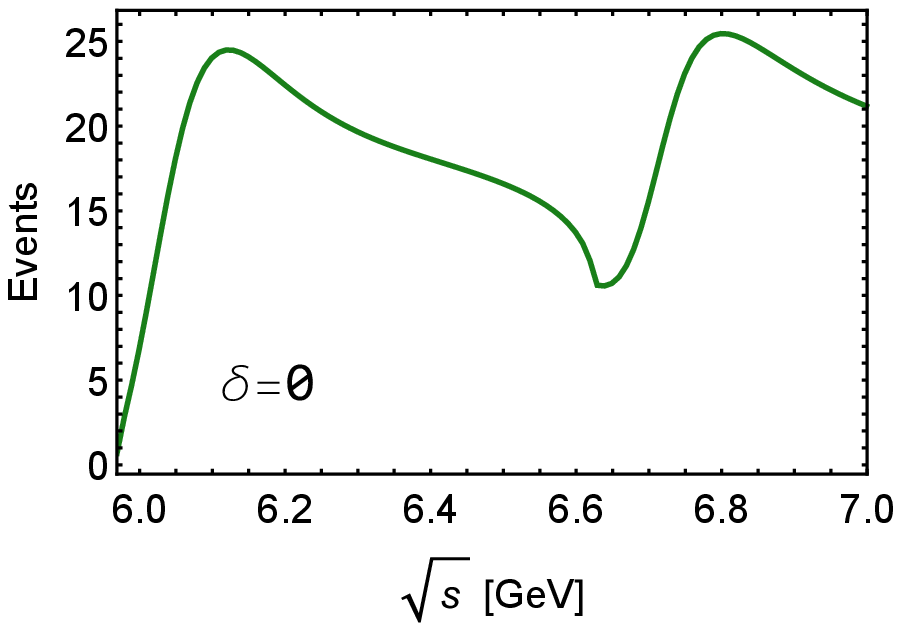}
}
\caption{The di-$\eta_{c}$ spectra in the coupled-channel model with different values for the phase angle $\delta$. Plots (a)-(h) correspond to $\delta= -\pi/4, \ -\pi/2, \ -3\pi/4, \ -\pi, \ , \ -\pi/4, \ \pi/2, \ 3\pi/4, \ 0$, respectively.}
\label{fig-phase-angle}
\end{figure}

In Fig.~\ref{fig-phase-angle} we present the results for the di-$\eta_c$ spectra with several different values for $\delta$, i.e. $\delta= -\pi/4, \ -\pi/2, \ -3\pi/4, \ -\pi, \  \pi/4, \ \pi/2, \ 3\pi/4, \ 0$. It shows that for all these phase angles, the threshold peak due to the low-mass pole at $(6041-i 170)$ MeV does not vary significantly. In contrast, the line shape near the $\eta_c$-$\eta_c(2S)$ threshold is very sensitive to the phase angle. This is understandable since the resonance pole, $(6711-i 141)$ MeV, is located at the fourth Riemann sheet which is close to the physical one. Therefore, the interference effects from the low-mass pole turn out to be significant. As shown in Figs.~\ref{fig-phase-angle} (e) and (f), the line shapes around the $\eta_c$-$\eta_c(2S)$ indicate apparent destructive interferences between these two poles and dip structures can be produced. It is also noticeable that the positions of the second enhancements in Figs.~\ref{fig-phase-angle} (d) and (h) are actually different due to the interfering effects.  We anticipate future experimental measurements should provide more stringent constraint on the phase parameter.

\section{Summary}

The observation of $X(6900)$ and the possible $X(6300)$ in the di-$J/\psi$ invariant mass spectrum at LHCb has provided great opportunities for probing the fully-heavy tetraquark systems. Although there have been a lot of efforts from the quark potential models advocated to the understanding of the fully-heavy tetraquark states, it is not clear why only a few structures have been seen in experiment while many states predicted by the potential quark model are absent from observation. In this study we show that the Pomeron exchange as the effective multi-soft-gluon exchange dynamics seems to play a crucial role in the near-threshold heavy quarkonium pair interactions. It is also a general feature for heavy-heavy systems which arises from the soft-gluon couplings to the heavy-flavor quarks. Following the study of $X(6900)$ and possible $X(6300)$ based on the Pomeron exchange mechanism, we extend this approach to the pseudoscalar charmonium pair system in a coupled-channel formalism. Although the lack of experimental information makes it impossible to constrain the di-$\eta_c$ production at LHC, we show that the near-threshold $\eta_c$-$\eta_c \ (\eta_c(2S))$ interactions via the Pomeron exchange can produce resonance poles with $J^{PC}=0^{++}$ and lead to non-trivial line shapes in the di-$\eta_c$ spectrum. This phenomenon can be studied at LHCb or in near-threshold di-$\eta_c$ production processes as a prediction for the Pomeron exchange mechanism. Nevertheless, as a general mechanism it can be searched for in other heavy quarkonium interactions in experiment. Further theoretical studies of its broad manifestations are also strongly recommended.

\begin{acknowledgments}
This work is supported, in part, by the National Natural Science Foundation of China (Grant No. 11521505), the DFG and NSFC funds to the Sino-German CRC 110 ``Symmetries and the Emergence of Structure in QCD'' (NSFC Grant No. 12070131001, DFG Project-ID 196253076), National Key Basic Research Program of China under Contract No. 2020YFA0406300, and Strategic Priority Research Program of Chinese Academy of Sciences (Grant No. XDB34030302).
\end{acknowledgments}

\bibliographystyle{unsrt}

\end{document}